\documentclass[runningheads]{llncs}
\usepackage{graphicx}
\usepackage{amsmath}
\usepackage{amssymb}
\usepackage{cleveref}
\usepackage{float}
\usepackage{tabularx}
\usepackage{booktabs}
\usepackage{multicol}
\usepackage{multirow}
\usepackage{siunitx}
\usepackage{rotating}

\begin{document}

\title{Contrast-Agnostic Groupwise Registration by Robust PCA for Quantitative Cardiac MRI}

\titlerunning{Contrast-Agnostic rPCA Groupwise Registration}

\author{Xinqi Li\inst{1} \and Yi Zhang\inst{1} \and Yidong Zhao\inst{1} \and Jan van Gemert\inst{2} \and Qian Tao\inst{1}
}
\authorrunning{X. Li et al.}

\institute{Department of Imaging Physics, Delft University of Technology \and 
Computer Vision Lab, Delft University of Technology}

\maketitle   
\begin{abstract}
Quantitative cardiac magnetic resonance imaging (MRI) is an increasingly important diagnostic tool for cardiovascular diseases. Yet, co-registration of all baseline images within the quantitative MRI sequence is essential for the accuracy and precision of quantitative maps. However, co-registering all baseline images from a quantitative cardiac MRI sequence remains a nontrivial task because of the simultaneous changes in intensity and contrast, in combination with cardiac and respiratory motion. To address the challenge, we propose a novel motion correction framework based on robust principle component analysis (rPCA) that decomposes quantitative cardiac MRI into low-rank and sparse components, and we integrate the groupwise CNN-based registration backbone within the rPCA framework. The low-rank component of rPCA corresponds to the quantitative mapping (i.e. limited degree of freedom in variation), while the sparse component corresponds to the residual motion, making it easier to formulate and solve the groupwise registration problem. We evaluated our proposed method on cardiac T1 mapping by the modified Look-Locker inversion recovery (MOLLI) sequence, both before and after the Gadolinium contrast agent administration. Our experiments showed that our method effectively improved registration performance over baseline methods without introducing rPCA, and reduced quantitative mapping error in both in-domain (pre-contrast MOLLI) and out-of-domain (post-contrast MOLLI) inference. The proposed rPCA framework is generic and can be integrated with other registration backbones.

\end{abstract}

\section{Introduction}

Quantitative cardiac MRI, such as T1 and T2 mapping \cite{messroghli2004modified}, is an increasingly important imaging modality to examine cardiovascular diseases~\cite{de2014cardiac}. However, the quality of quantitative mapping is negatively affected by respiratory and cardiac motion during the MR acquisition procedure~\cite{xue2012motion}. Such motion leads to misalignment of tissue across baseline images, resulting in deteriorated accuracy and precision of the final quantitative mapping~\cite{kellman2013t1}. To improve the quality of quantitative cardiac MRI, motion correction by deformable image registration is an essential part of the post-processing pipeline~\cite{ashburner2007fast,chen2021deep,makela2002review}. 

Conventionally, deformable image registration is implemented in a pairwise fashion: each time two images are registered, with one designated as fixed and one moving. However, for quantitative cardiac MRI, the number of  images is highly variable (ranging from 3 to $>$ 20) depending on the specific sequence. This makes pairwise registration not a natural option, as the ``best" fixed image is hard to define. Moreover, registration error easily propagates across the baseline images, given all pairwise registration steps are independently performed. The alternative approach of \emph{groupwise image registration}, which registers all baseline images simultaneously, was proposed for quantitative MRI motion correction~\cite{huizinga2016pca,tao2018robust,feng2016liver,geng2009implicit}. Groupwise registration promises improved robustness across a sequence of images by optimizing a global metric which promotes co-registration of \emph{all} frames, including those with extremely poor contrast and hence difficult to register in a pairwise fashion. Groupwise image registration can be divided into two paradigms: classical iterative optimization methods that are relatively slow~\cite{hamy2014respiratory,huizinga2016pca,tao2018robust,feng2016liver,guyader2018groupwise,li2022motion} and deep-learning-based methods that promise fast inference~\cite{ahmad2019deep,che2019deep,fechter2020one,zhang2021groupregnet,gonzales2021moconet}. 

A special challenge in motion correction for quantitative cardiac MRI is that the change in image contrast and intensity can vary drastically across baseline images, completely agnostic to the image registration pipeline~\cite{xue2012motion}. The pattern of variation, which is determined by the underlying MR signal model, differs per quantitative sequences; even with the same signal model, the contrast is still dependent on the exact scheme of acquisition, which differs again among MRI machines. This makes it difficult to design a consistently reliable registration metric for optimization. Conventional registration metrics, such as NCC and NMI, can still be sensitive to agnostic contrast changes and fail~\cite{brudfors2020groupwise,de2020mutual,klein2009evaluation}. Therefore, finding a robust registration metric in the face of agnostic contrast changes is of great interest.

Furthermore, we observed that an under-studied phenomenon is the degenerated solution of groupwise registration, in the format of ghosting artefacts or pixel collapse~\cite{de2020mutual}. These degenerated solutions lead to an optimal metric, but are implausible because they violate the anatomical consistency. In this paper, we will further investigate the susceptibility of NCC and NMI to such artifacts.

In this work, we set out to tackle the agnostic contrast change in quantitative cardiac MRI by designing a novel registration framework, which integrates robust PCA (rPCA)~\cite{candes2011robust} with state-of-the-art image registration backbones. Our rationale of introducing rPCA is as follows: firstly, the signal model, which is typically governed by physics principles, has a limited degree of freedom~\cite{chow2014saturation,messroghli2004modified}, underlying the \emph{low-rank} component of rPCA. Secondly, the motion of quantitative cardiac MRI is \emph{sparse} in the sense that it is often concentrated around the heart, induced by non-ideal breath-hold and heart rate variability, while the background, e.g., rib cage and lung, stay largely static. Decomposition of the two components creates ease for registration algorithms. In this paper, we propose to integrate rPCA with the state-of-the-art deep-learning groupwise registration method~\cite{zhang2021groupregnet} for fast, reliable motion correction of quantitative cardiac MRI. Our main contributions are:
\begin{enumerate}
    \item We propose a novel groupwise image registration framework, which is, to the best of our knowledge, the first attempt to utilize rPCA in groupwise registration with a deep learning backbone. This generic framework can be integrated with any existing registration methods, either classical optimization or modern deep learning methods.

    \item We evaluated and demonstrated the generalizability of our contrast-agnostic method on out-of-domain quantitative MRI sequences.

    \item We further investigated the fitness of two popular metrics, NCC and NMI, for groupwise registration. We showed empirically that NCC could give rise to registration artefacts, leading to unwanted anatomical deformation. 
\end{enumerate}
\section{Methods}
\subsection{Problem Formulation}
\begin{figure}[t]
\centering
\includegraphics[width=0.85\textwidth]{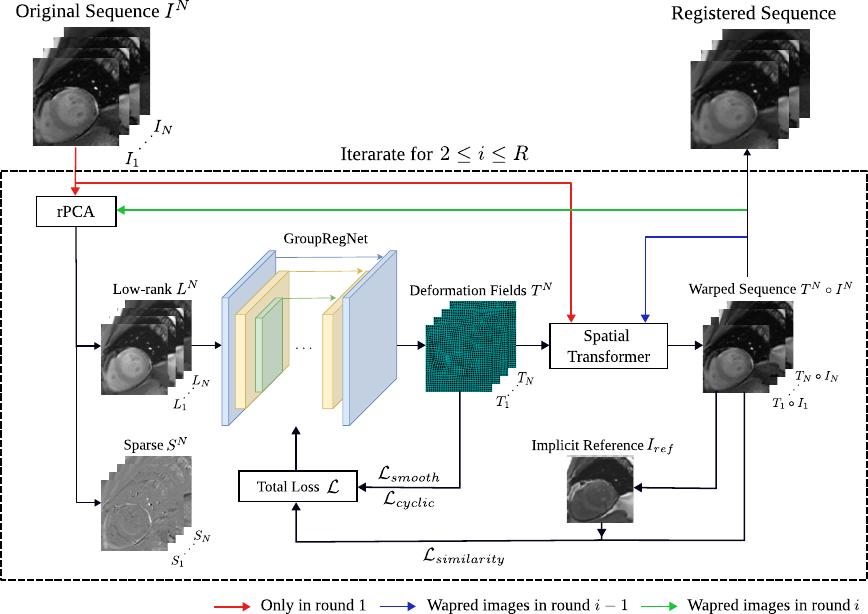}
\caption{Overview of the proposed framework for contrast-agnostic registration. The dotted rectangle denotes the iterative registration pipeline that progressively corrects motion from round $2$ to maximal round $R$.} \label{workflow}
\end{figure}
Given a sequence of baseline images $I^N = \{I_i \in \mathbb{R}^{H \times W}|i=1, ..., N\}$, the goal of groupwise registration is to align all $I_i$ into one common coordinate system by obtaining a set of deformation fields $T^N = \{T_i \in \mathbb{R}^{2 \times H \times W}; i=1, ..., N\}$. An implicit reference $I_{ref} = \frac{1}{N}\sum_{n=1}^N(T_n \circ I_n)$ is generated for groupwise registration. Therefore, each $T_i$ should align the anatomical structures in $I_i$ to those in the implicit reference $I_{ref}$. 

The proposed rPCA framework is as follows: rPCA first decomposes the input baseline images $I^N$ into the low-rank matrix $L^N$ and sparse matrix $S^N$. The low-rank component is fed to the deep learning backbone to learn the deformation field $T^N$. Then $T^N$ is applied to the input $I^N$ to obtain warped images $T^N \circ I^N$, which serves as the input of the next iteration of rPCA, until the maximal iteration number is reached. The framework then progressively corrects the motion in the original input, but for each iteration, rPCA enables us to work only on the low-rank part, which is easier to register than the original input. This rationale will be revisited later in the paper. The diagram of our proposed framework is shown in \Cref{workflow}. 
\subsection{Robust Principal Component Analysis}
Robust principal component analysis (rPCA)~\cite{candes2011robust}, as its name suggests, is a robust version of PCA for matrix decomposition: For a given data matrix $M$, where in our case $M$ is the matrix of vectorized grouped images $I^N$, the rPCA decomposes $M \in \mathbb{R}^{m \times n}$ into the sum of a low-rank matrix $L$ and a sparse matrix $S$ via solving the following optimization problem:

\begin{equation}
\text { minimize }  \|L\|_*+\lambda\|S\|_1, \text { subject to } L+S=M,
\end{equation}
where $\| \cdot \|_*$ denotes the nuclear norm, $\| \cdot \|_1$ denotes the $l_1$ norm, and $\lambda$ is a hyperparameter trading off the two components, which is often set by default as $\lambda = 1/{\sqrt{\max{(m,n)}}}$. Such optimization problems can be solved by well-established algorithms, such as proximal gradient descent methods~\cite{feng2013online}.
\begin{figure}[t]
\includegraphics[width=\textwidth]{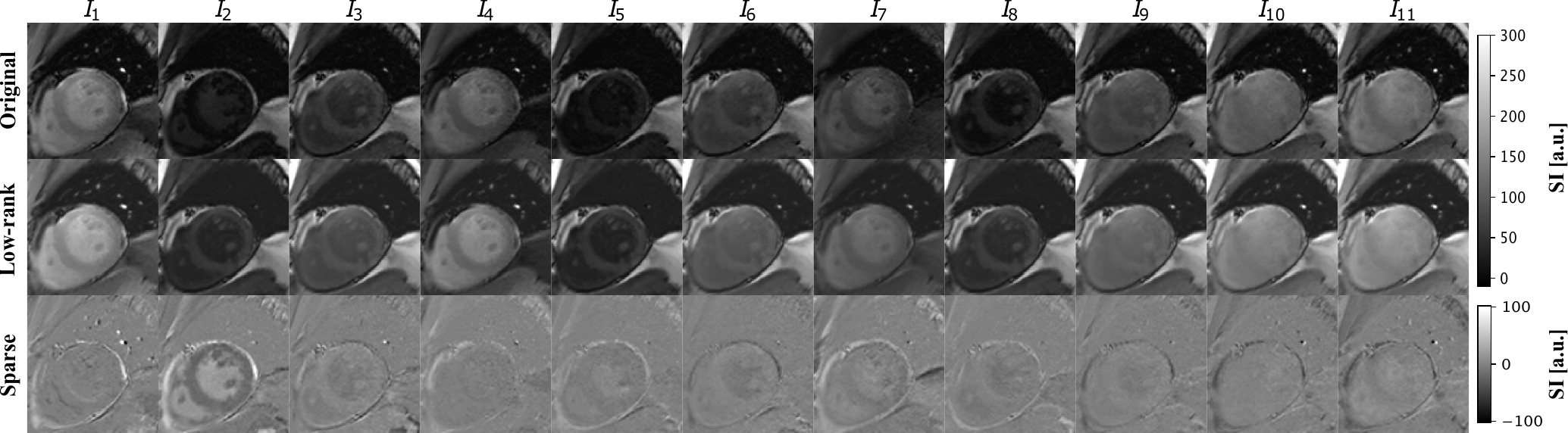}
\caption{Decomposition of pre-contrast MOLLI cardiac time-series images using rPCA. Each MOLLI sequence consists of 11 pre-contrast time frames in our setting. SI denotes signal intensity. The intensity inconsistency of the sequence is mitigated in the low-rank component of rPCA.} \label{rPCA_illust}
\end{figure}

An illustration of rPCA on pre-contrast cardiac MRI is shown in \Cref{rPCA_illust}. It can be seen that the sparse matrix captures abrupt changes in baseline images, either of contrast (such as in $I_2$) or motion (such as in $I_7$), which are usually more difficult to handle by registration algorithms. The low-rank matrix, in contrast, consists of the components in baseline images which have relatively lower variations and are easier to align. Note that the amplitude of the sparse component is much lower than that of the low-rank component, the latter capturing the majority of information in the original matrix.

\subsection{Loss Functions}
The optimization problem for finding the deformable mapping $T^N$ can be formulated as follows:

\begin{equation}
T^N = \mathop{\arg \min}\limits_{T^N} \mathcal{L}_\mathrm{similarity} + \lambda_0 \mathcal{L}_\mathrm{smooth} + \lambda_1 \mathcal{L}_\mathrm{cyclic},
\label{eq_loss}
\end{equation}
where $\mathcal{L}_\mathrm{similarity}$, $\mathcal{L}_\mathrm{smooth}$, and $\mathcal{L}_\mathrm{cyclic}$ denote similarity function, smoothness regularization, and cyclic consistency, with weight parameters $\lambda_0$ and $\lambda_1$.

\noindent\textbf{Similarity Functions:} We employed the normalized mutual information (NMI) to measure the similarity between the input images $I^N$ to the warped images $T^N \circ I^N$, which can measure alignment in the face of contrast changes~\cite{de2020mutual}. The NMI between two images is defined as:
\begin{equation}
NMI(I_1, I_2) = \frac{2 MI(I_1, I_2)}{H(I_1) + H(I_2)},
\label{eq1}
\end{equation}
where $MI(I_1,I_2)$ denotes the mutual information between $I_1$ and $I_2$, $H(I_1)$ is the entropy of image $I_1$, and $H(I_2)$ for image $I_2$, respectively. For groupwise registration, the similarity loss $L_{similarity}$ is then defined as:
\begin{equation}
\mathcal{L}_\mathrm{similarity} = - \frac{1}{N}\sum_{n=1}^N NMI(T_n \circ I_n, I_{ref}).
\label{eq2}
\end{equation}

Another popular similarity loss is also considered and discussed, which is the local normalized cross-correlation (NCC)~\cite{avants2008symmetric}, defined as
\begin{equation}
NCC(I_1, I_2) = \frac{1}{H\times W}\sum_{i, j \in H, W} \frac{\sum_{x \in \Omega}(I_1(x) - \bar{I_1}(i, j))(I_2(x) - \bar{I_2}(i, j))}{\sqrt{\hat{I}_1(i ,j)\hat{I}_2(i, j)}},
\label{eq2}
\end{equation}
where $H$ and $W$ corresponds to the height and width of the image, $\Omega$ indicates the neighborhood voxels around the voxel at position $(i, j)$ and $\bar{I}(i, j)$ and $\hat{I}(i ,j)$ denote the local mean and variance.

\noindent\textbf{Smoothness Regularization:} The smoothness of the deformation field is regularized through B-spline registration~\cite{rueckert1999nonrigid}. We adopted B-spline because it can prevent the image from folding and inherently lead to smooth deformation fields: 
\begin{equation}
\mathcal{L}_{\text {smooth }}=\frac{1}{H \times W} \sum_{n=1}^{N}\int_0^H \int_0^W\left[\left(\frac{\partial^2 \hat{T}_n}{\partial x^2}\right)^2+\left(\frac{\partial^2 \hat{T}_n}{\partial y^2}\right)^2 +2\left(\frac{\partial^2 \hat{T}_n}{\partial x y}\right)^2\right]d x d y,
\end{equation}
where $\hat{T}_n = T_n + \sum_{l=0}^k \sum_{m=0}^k B_l(u) B_m(v) \phi_{i+l, j+m}$, and $B_l$ is the $l-$th B-spline basis function, $k$ is the order of B-spline, and $\phi_{i,j}$ denotes the control points with uniform space across the image. B-Spline control points will affect the surrounding deformation fields based on the kernel function.

\noindent\textbf{Cyclic Consistency:} For groupwise registration, the cyclic consistent regularization keeps the estimated implicit reference at the center of all baseline images in the manifold by minimizing the deformation field to the implicit reference \cite{zhang2021groupregnet}:
\begin{equation}
\mathcal{L}_\mathrm{cyclic}(T^N) = \sqrt{\frac{1}{2(H\times W)}\sum_{i, j \in H, W}\left(\sum_n T_{n}(i, j)\right)^2},
\label{eq2}
\end{equation}
where $T_n(i,j)$ denotes the value of $T_n$ at coordinate $(i,j)$. This term prevents the degenerated solution where textures in all images collapse.

\subsection{CNN-based Neural Network Architecture}
The convolution neural network architecture follows that of the VoxelMorph \cite{balakrishnan2019voxelmorph}, and GroupRegNet \cite{zhang2021groupregnet}, based on the UNet \cite{ronneberger2015u} architecture consisting of encoding and decoding layers with skip connection. Both encoder and decoder use convolutional blocks consisting of a 2D convolution and a Leaky ReLU activation function. The encoder captures the hierarchical features of the input images with multiple convolution blocks. The number of decoder layers was controlled by the B-spline kernel size $k$~\cite{ronneberger2015u}. The larger kernel size indicates less decode layers, which makes the deformation field more homogeneous. This enables the coarse-to-fine representation of the two-channel deformation field. The final deformation field is computed by B-spline free form deformation (FFD) transformation model \cite{rueckert1999nonrigid} based on the decoder output.

\subsection{Evaluation Methods} 
\noindent \textbf{T1 fitting error:} In this paper, we used myocardial T1 mapping by the modified Look Locker inversion recovery sequence (MOLLI), one of the most widely used mapping modalities in clinical practice~\cite{messroghli2004modified}. T1 mapping follows a three-parameter model, expressed by
\begin{equation}
y(T_I) = A - B e^{-T_I/T_1^*},
\label{eq: T1}
\end{equation}
where $y$ denotes the signal intensity, $T_I$ denotes the inversion time for acquisition of each baseline image, and $A$, $B$, and $T1^*$ are parameters to be estimated. Since motion correction leads to a better fitting of this MR physics model at each pixel, here we measure the performance through the T1 mapping within the ROI (myocardium and left ventricle) and the standard deviation (SD) error~\cite{kellman2013t1} as an indication of the fitting error. A lower SD error indicates better motion correction. We used both the native (pre-contrast) T1 mapping and post-contrast T1 mapping sequences (after Gadolinium administration). To test the generalizability of our framework, we trained our NN exclusively on pre-contrast T1 mapping, while testing it on both pre-contrast (in-domain) and post-contrast (out-of-domain) sequences. 

\noindent \textbf{Dissimilarity metrics $\mathcal{D}_{PCA}$:} We evaluate the warped images using $\mathcal{D}_{PCA}$, the ratio of the top-$K$ eigenvalues to the sum of eigenvalues of the correlation matrix \cite{huizinga2016pca}. The higher the ratio, the better the performance of registration.

\noindent \textbf{Baseline methods:} We compared our proposed framework with two methods: (1) the conventional groupwise method, Elastix-PCA~\cite{huizinga2016pca}, and (2) the groupwise registration method~\cite{zhang2021groupregnet} without rPCA, denoted by \textit{GroupRegNet$^*$}, whichfollows~\cite{zhang2021groupregnet}, but using NMI as the optimization metric. We also performed experiments on \textit{GroupRegNet$^*$} using the NCC metric as in the original work and compared the results with NMI.
\section{Experiments and Results}

\textbf{Dataset:} We used a cardiac MRI dataset including 48 subject, with both pre-contrast and post-contrast MOLLI sequences (Philips 3.0T). Each subject had 1 to 3 slices acquired at the base, mid-ventricular, and apex levels. In total 120 pre-contrast and 120 post-contrast MOLLI sequences were included. All images were resampled to a $224 \times 224 \times 11$ grid with \qty{1}{\cubic\milli\meter} isotropic resolution and then cropped to $112 \times 112 \times 11$ at the center. The training comprised 100 random images from only the \emph{pre-contrast} MOLLI sequences. The rest 20 pre-contrast MOLLI sequences and their corresponding post-contrast sequences, in total 40, formed the test set. We note here that the pre-contrast sequences are the \emph{in-domain} test data, while the post-contrast sequences are the \emph{out-of-domain} test data, given their contrast changes follow a different pattern governed by much higher relaxation rate due to the contrast agent (e.g. lower T1).
\begin{table*}[ht!]
     \caption{Experiment results on T1 mapping. We compare T1 SD and $\mathcal{D}_{PCA}(K=1)$ before and after registration. A higher $\mathcal{D}_{PCA}(K=1)$ indicates larger power in the princinple components thus better alignment. The SD measures the $T1$ fitting error within the ROI. Lower SD indicates lower fitting error thus better alignment. Our method (w/ rPCA) outperforms the GroupRegNet$^*$ on both pre-contrast and post-contrast data in terms of both SD and $\mathcal{D}_{PCA}$. The bold values demonstrates the best result and underlined values is the second highest performance for each metric.}
    \centering
        
    \begin{tabular}{p{1.5cm} p{2.8cm}<{\centering} m{2.2cm}<{\centering} m{2.2cm}<{\centering} m{2.2cm}<{\centering} m{2.3cm}<{\centering} m{1.2cm}<{\centering}}
        \toprule
        Modality & Method & SD($\unit{\ms}$)$\downarrow$  & $\mathcal{D}_{PCA}\%$$\uparrow$ & Time($\unit{\s}$) \\
        \midrule
        \multirow{3}{*}{Pre-GD} &
        Elastix-PCA  & \underline{54.5$\pm$21.7} & 93.9 & $ \approx 600$\\

        & GroupRegNet$^*$  & 55.4$\pm$21.4 & 94.0 & 1.28 \\
        & Ours (w/ rPCA)&  \textbf{53.9$\pm$21.9} & \textbf{94.4} & $7.11$\\
    \midrule    
        \multirow{3}{*}{Post-GD} &Elastix-PCA &  \textbf{21.5$\pm$16.1} & 91.7 & $ \approx 600$\\
        & GroupRegNet$^*$ & 24.5$\pm$13.3 & 81.5 & 1.28 \\
        & Ours (w/ rPCA)& \underline{21.8$\pm$12.5} & \textbf{92.1} & $7.11$\\
        \bottomrule

        \end{tabular}

    \label{tab-label-2}

\end{table*}

\begin{figure}[ht!]
    \centering
    \includegraphics[width=0.99\textwidth]{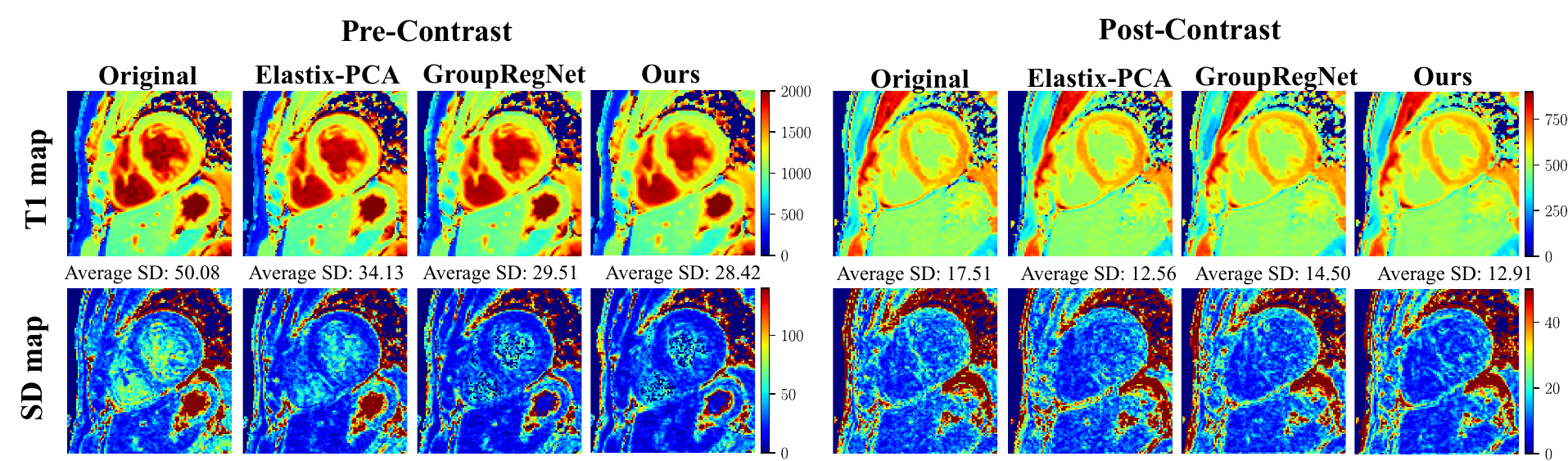}
    \caption{Representative figures of the quantitative CMR. The figure demonstrated the T1 map (top row) and SD map (bottom row) of the same subject on both pre- and post-contrast. We compared our proposed method with conventional method (Elastix-PCA) and deep learning based GroupRegNet, and the reported average SD within the ROI indicated that our method outperformed others.}
    \label{fig:representative}
\end{figure}
\noindent\textbf{Implementation Details:}
Robust PCA was implemented with the GoDec algorithm~\cite{zhou2011godec}. In each round, the rank of $L$ was set to be half of the sequence length, which was $5$ in our case. Empirically we applied a decoder with 4 layers. In this case, the decoder included 2 convolution blocks and the output deformation field was $31 \times 31 \times 11 \times 2$. The final deformation field was transformed to $112 \times 112 \times 11 \times 2$ using B-spline FFD. The smooth regulation's weight $\lambda_0$ is set to $0.001$ and cyclic regulation's weight $\lambda_1$ is set to $0.01$ empirically.

\noindent \textbf{Choice of Similarity Functions:}
Two similarity functions, NCC and NMI, are evaluated. We observed that NCC loss led to undesirable deformation as well as altered distribution of the T1 values (details in Supplementary). As suggested in~\cite{de2020mutual}, NCC naturally favors homogeneous distribution of pixel intensities and lead to over-smooth myocardium textures that fail the purpose of quantitative mapping, while NMI maintained the shape and texture of the ROI.\\

\noindent \textbf{Results:}
The quantitative results of registration and quantitative mapping are shown in \Cref{tab-label-2} and two representative figures are shown in \Cref{fig:representative}. Note that to demonstrate the generalizability of the learned model, we train the model \textbf{only} on pre-contrast data (denoted as Pre-Gd) and tested on both pre- and post-contrast data. Our method performs best on pre-contrast datasets according to SD within the ROI and outperforms the GroupRegNet$^*$ on post-contrast data. Elastix-PCA gives slightly better performance on post-contrast data because the optimization is per datasets (no training and inference). However, it takes around 10 minutes for each subject, which is much slower compared to our method, with an average inference time of \qty{7.1}{\s} per sequence.
\section{Conclusion}
In conclusion, we proposed a novel rPCA framework for robust motion correction of quantitative cardiac MRI. We aim for robust performance despite agnostic image contrast changes, which are typical of quantitative MRI. We showed that the introduction of rPCA, which separates low-rank and sparse components of baseline images, led to improved registration performance and facilitated the generalization of the trained network on out-of-domain data. 

In addition, our work also compared the two commonly used metrics for groupwise registration, namely, NCC and NMI, and showed that NCC might give rise to potential loss-specific artifacts in heart anatomy and quantitative mapping. Future investigations are warranted to focus not only on the performance of image registration but also on the fidelity of quantitative mapping.

\bibliographystyle{splncs04}
\bibliography{ref}

\newpage
\appendix
\section{Supplementary}

\begin{figure}
    \centering
    \includegraphics[width=\linewidth]{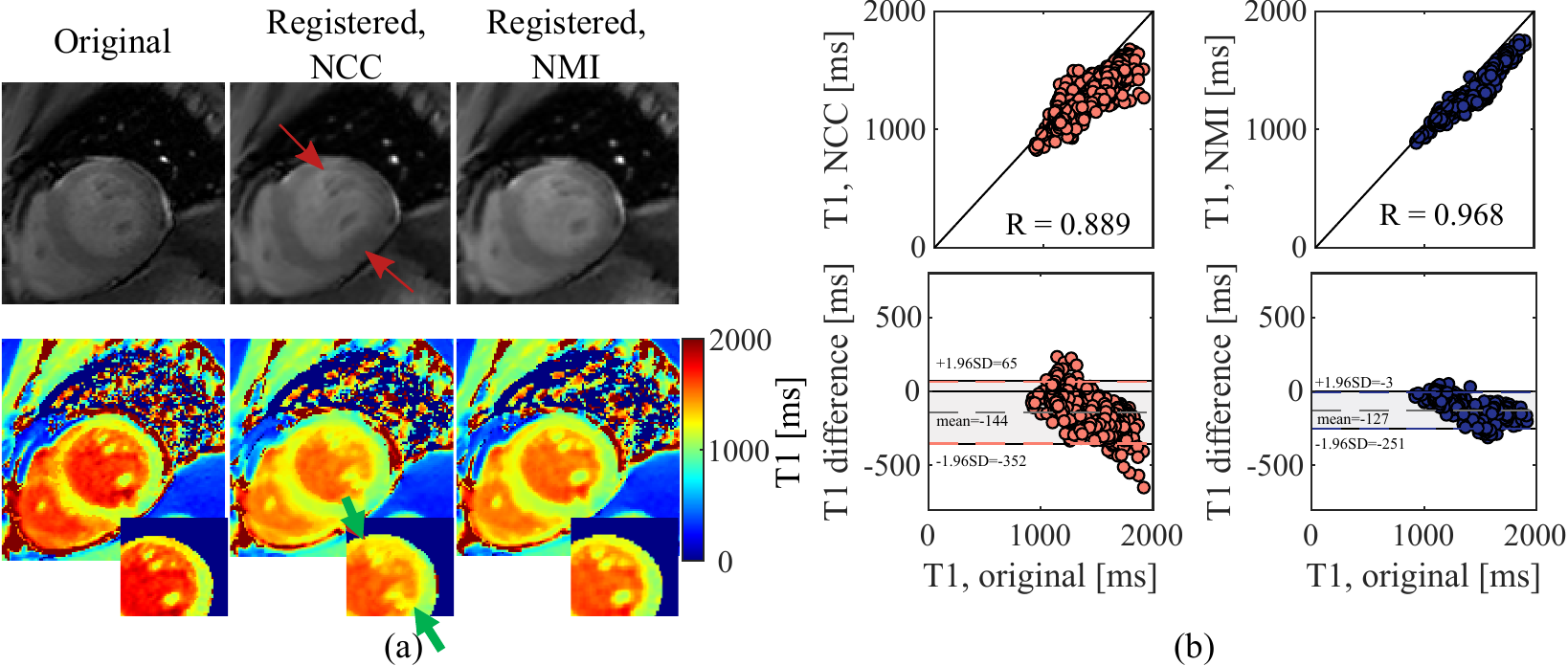}
    \caption{We compared two metrics (normalized cross correlation and normalized mutual information) by training on the same training data and checked the result after registration and parametric fitting. Left panel (a) shows an exemplar case of the original and registered images using NCC (middle) and NMI (right). The second row shows the resulting T1 map. Right panel (b) shows the correlation and Bland-Altman plot of the quantitative T1 estimation within the myocardium ROI. Red arrows in (a, top row) point to the potential deformation artifact of the NCC-guided registration, in the form of implausible anatomical deformation and biased T1 estimation (b). Green arrows in (a, bottom row) indicated the corresponding changes of anatomical pattern in T1 maps.}
    \label{fig:enter-label}
\end{figure}

\end{document}